\documentclass[12pt]{article}
\usepackage{amsmath}
\usepackage[margin=2.0cm]{geometry}
\usepackage{epsfig}
\usepackage{amsfonts}
\usepackage{pifont}
\usepackage{amssymb}
\usepackage{float}
\usepackage{xcolor}
\usepackage{titlesec}
\usepackage{fancyhdr}
\usepackage{cite}
\usepackage{algorithm}
\usepackage{algpseudocode}
\usepackage{hyperref} 

\hypersetup{
  colorlinks=true,
  linkcolor=magenta,
  citecolor=blue,
  urlcolor=cyan,
  pdftitle={Dawara \& Viswanathan},
  pdfpagemode=FullScreen,
}

\usepackage{setspace}
\usepackage{authblk}
\usepackage{cite}
\newcommand{\runningtitle}{} 
\newcommand{\authorname}{Vineet Dawara and Koushik Viswanathan\thanks{koushik@iisc.ac.in}} 
\newcommand{\suppMat}{\textbf{Supplementary Material}}
\newcommand{\suppref}[1]{{~\color{magenta}{#1}}}

\titleformat{\section}
{\normalfont\bfseries\scshape}{\thesection}{1em}{}

\titleformat{\subsection}
{\normalfont\bfseries\scshape}{\thesubsection}{1em}{}

\titleformat{\subsubsection}
{\normalfont\slshape}{\thesubsubsection}{1em}{}

\titleformat{\title}{\normalfont\bfseries}{\thesection}{1em}{}

\title{\vspace{-3em} \large \bfseries Dynamic fragmentation of residually stressed solids: From microscopic instabilities to universal scaling}
\author{\normalsize \authorname}
\affil{\vspace{-1em}  \normalsize \textsl{Department of Mechanical Engineering, Indian Institute of Science, Bangalore}}

\date{\normalsize \textsc{\today}}
\begin{document}
\maketitle
\thispagestyle{plain}
\hrulefill

\begin{abstract}
The dynamic fragmentation of residually stressed solids involves a complex interplay between stored elastic energy, stress wave propagation, and crack instabilities. In this work, we investigate the fracture mechanics of chemically toughened glass through high-velocity projectile impact experiments and a novel micromechanical network model. We rigorously incorporate residual stress into the discrete lattice framework via a prescribed inelastic strain (eigenstrain) distribution, formulated as equivalent body and surface forces to ensure mesh-independent fracture paths. Our experiments and simulations demonstrate that while the fracture topology shifts from coarse to fine with increasing impact energy, the cumulative fragment size distribution consistently follows an exponential decay. Crucially, we reveal a universal scaling law: fragment size distributions from diverse loading conditions and stress profiles collapse onto a single master curve when normalized by the mean fragment area. Furthermore, the model elucidates the determinants of fragmentation, showing that the resulting fragment size is governed not only by the magnitude of residual stress but also by the steepness of the stress gradient. At the microscopic scale, we identify a mechanism for dynamic instability where non-sequential bond breaking ahead of the crack tip leads to apparent local crack speeds exceeding the Rayleigh wave speed ($c_r$). These arrested micro-branches, analogous to the Burridge-Andrews mechanism, provide a physical explanation for the \lq tongue-like\rq\ features and hackle zones observed in post-mortem fractography.
\end{abstract}

\hrule

\section{Introduction}
\label{sec:intro}

Residual stresses can often be deliberately introduced in solids to enhance their structural performance, the most notable example being that of toughened glass \cite{tandon2015fracture}. This is often achieved either by thermal processing, where rapid cooling creates differential contraction between the surface and the interior \cite{narayanaswamy1969calculation, nielsen2009fracture, pourmoghaddam2018experimental}, or by chemical ion exchange, where larger ions (e.g., K$^+$) replace smaller ions (e.g., Na$^+$) to create internal strain \cite{garfinkel1970ion, arakawa2016ultrasonic}. Both processes introduce incompatible deformations (eigenstrains), resulting in a compressive stress near the surface and a tensile core. This configuration is metastable: while the compressive stress suppresses the growth of surface flaws, tension in the interior promotes catastrophic fragmentation if a crack penetrates through. Due to this dual benefit of strength and safety---shattering produces small, non-hazardous fragments---toughened glass has been integral to industrial applications for over a century \cite{gardon1980thermal}, ranging from architectural facades to automotive components \cite{dix2022digital, efferz2024photoelastic}.

While direct measurement of the residual stress field is experimentally challenging \cite{arakawa2016ultrasonic, tandon1990residual}, correlations between its qualitative features and the corresponding processing parameters have been well studied and tabulated \cite{efferz2024photoelastic}. This is commonly done using experimental stress analysis (e.g., photoelasticity) necessarily supplemented by 3D finite-element simulations \cite{pourmoghaddam2016numerical, pourmoghaddam2018finite}. Such processing-property relationships are important from a practical viewpoint, however, they do not shed much light on the underlying mechanics \cite{narayanaswamy1969calculation, soules1987finite, nielsen2010finite, nielsen2010simulation}. In fact, the coupling between the residual stress field and the dynamic topology of the resulting fracture and fragmentation pattern remains a largely open question.

The chaotic shattering process, first documented in the historical context of Prince Rupert's drops by Robert Hooke \cite{hooke1665micrographia}, is the result of multiple, simultaneous crack bifurcation events \cite{aben2016extraordinary}. The topology of this network, including the number of fragments, as well as their characteristic size, is closely linked to the underlying residual stress field. Experimental studies suggest that the fragmentation of toughened glass follows an exponential size distribution \cite{kooij2021explosive}, whereas the fragmentation of ordinary glass (or, indeed, any prototypical brittle solid under dynamic loading) typically exhibits a scale-free power-law distribution \cite{grady1985geometric, grady2008fragment}. This discrepancy implies the emergence of a characteristic length scale in residually stressed solids. The physical mechanism dictating this length scale, and its quantitative link to the stored elastic energy density, remains to be fully elucidated. The reason for this lacuna is that fragmentation is inherently a multi-crack phenomenon, wherein interactions between closely spaced cracks is largely responsible for the resulting facture network \cite{kooij2021explosive}.

Early explanations of fracture patterns and length scales leveraged Linear Elastic Fracture Mechanics (LEFM) to analyze a single propagating crack. Here stored residual elastic energy contributes to the stress intensity factor ahead of the moving tip---once a critical intensity is reached, the crack bifurcates \cite{barsom1968fracture, gulati1997frangibility}. Such simple energetics-based arguments can relate fragment size to bulk residual stress assuming idealized fragment shapes, e.g., hexagonal geometries based on Yoffe's theory \cite{yoffe1951lxxv}. Understandably, an approach like this cannot handle multiple cracks, making predictions predominantly qualitative in nature. At the other end of the spectrum are purely geometric or statistical models that postulate rules-based crack branching \cite{grady1985geometric}. While these models can predict the statistics of complete networks, their rules often lack a clear micromechanical basis, making them somewhat divorced from experimental data \cite{clemmer2022critical}. 

The ideal framework would therefore bridge this gap: it must incorporate mechanics-based criteria at the crack tip to mediate motion and bifurcation, while simultaneously handling an ensemble of interacting cracks within a pre-stressed environment. In this work, we realize such a framework by investigating the mechanics of fragmentation through combined projectile impact experiments and network-based numerical simulations. We conduct controlled impact tests using a gas gun based experimental framework to obtain fracture and fragmentation patterns accompanying residual stress fields. To deconstruct the microscopic fast crack dynamics inaccessible in the experiment, we present a micro-mechanical network model that incorporates heterogeneity arising from incompatible strains. While recent attempts have simulated dynamic fragmentation using modified finite element models \cite{vocialta2018numerical}, particle discretization schemes \cite{hirobe2021mathematical}, or peridynamics \cite{stewart2023modeling}, our model provides a computationally efficient framework that naturally handles material discontinuities. It implicitly captures multiple crack nucleation events and resolves the microscopic interactions between propagating cracks and the internal stress field. We demonstrate that this model reproduces the transition to an exponential distribution and isolates key factors—specifically projectile velocity and residual stress magnitude—to characterize the microscopic mechanisms governing crack branching.

The manuscript is organized as follows. Sec.~\ref{sec:methods} describes the methodology, outlining the experimental projectile–glass configuration and presenting the formulation of the spring-type micro-mechanical model with incompatible strains. Sec.~\ref{sec:results} presents our findings, comparing experimental and numerical results on fracture patterns and size distributions. We then investigate the influence of varying residual stress profiles and examine the microscopic mechanisms of crack branching. Finally, Sec.~\ref{sec:discussion} discusses the implications of the network model for fracture mechanics, followed by concluding remarks in Sec.~\ref{sec:conclusions}.

\section{Methods}\label{sec:methods}

This section details two methods used in the present manuscript: an experimental configuration, wherein a gas-gun-propelled dart penetrates the compressively stressed surface layer of toughened glass, and a network-based numerical formulation used to simulate the resulting fragmentation pattern.

\subsection{Experimental projectile-glass impact configuration}\label{sec:glass_exp}

\begin{figure}[h!]
	\centering
	\includegraphics[width = \columnwidth]{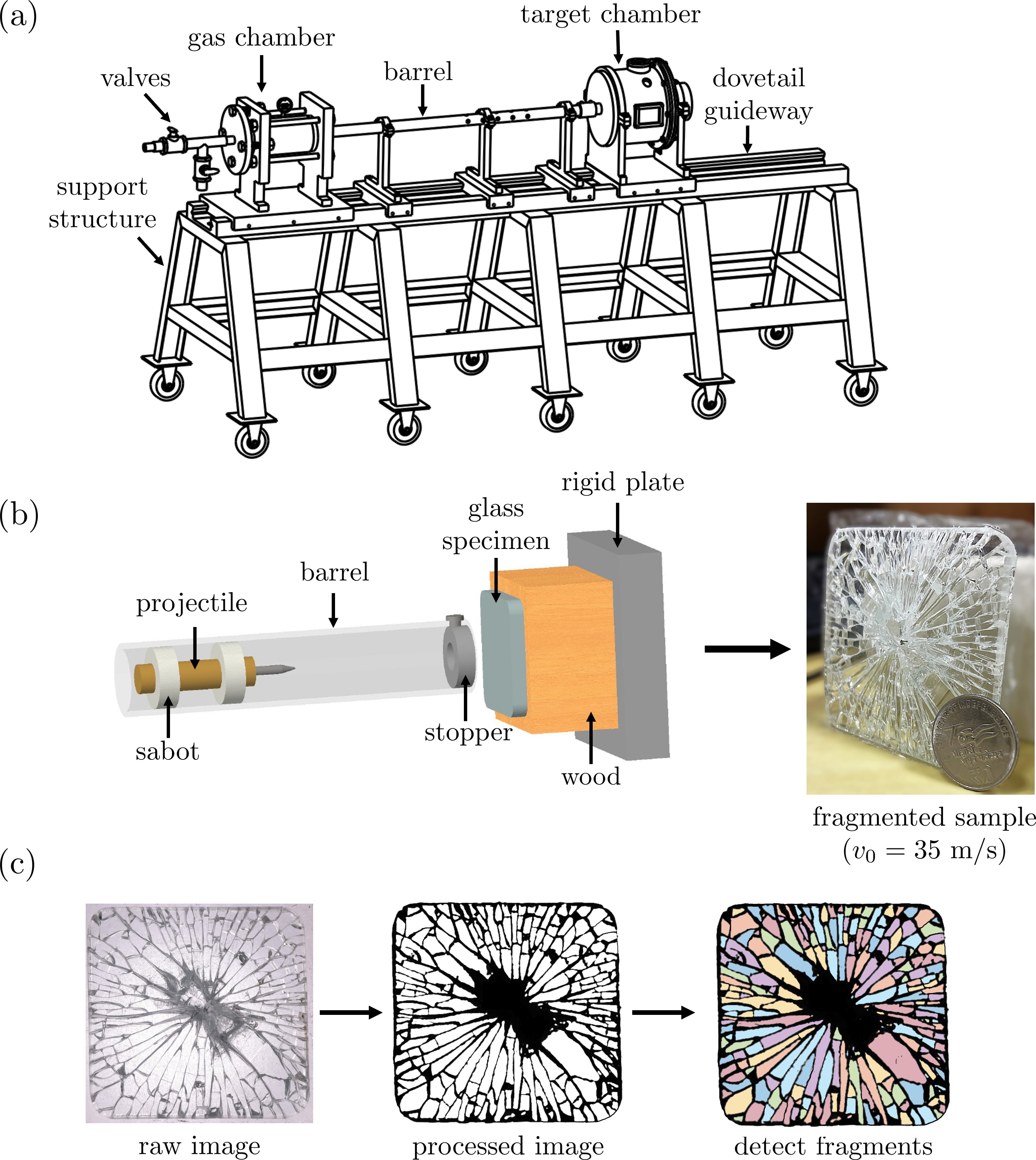}
	\caption{(a) Projectile geometry and (b) projectile-glass impact configuration, and (c) image processing of the fragmented sample for fragment detection.}
	\label{fig:exp_config}
\end{figure}

Impact experiments were conducted on commercially procured (Y.A. Industrial supplier, Bangalore) chemically toughened glass samples (60 mm $\times$ 60 mm $\times$ 5 mm) with a 5 mm fillet radius at the corners. The impact was generated using a pneumatic gas gun with a 21 mm barrel diameter that is described in detail elsewhere \cite{dawara2023design}. The setup comprises a gas chamber, a barrel, and a target chamber as shown in Fig.~\ref{fig:exp_config}(a). The gas chamber is pressurized with compressed air via a regulating valve. Operating conditions—specifically the initial chamber pressure and projectile mass—were determined using a calibration curve relating these parameters to resulting projectile speed. The velocity was independently measured using two infrared (IR) sensors positioned 100 mm apart on the barrel and recorded via a microcontroller. A detailed design of the gas gun and operating procedures are provided in Ref.~\cite{dawara2023design}.

The projectile is a pointed cylindrical stainless steel dart (2.4 mm diameter) with a $43^\circ$ tip angle, see Fig.~\ref{fig:exp_config}(b). The rear end is attached to an 8.25 mm thick rod mounted within Teflon sabots (21 mm outer diameter) to align the projectile exactly with the barrel axis. The sharp tip ensures fracture initiates precisely at the impact point with negligible flexural contribution.

The total projectile mass includes both the dart and the sabot. To achieve the desired impact velocities ($v_0$) of 20 m/s and 35 m/s, the sabot length was adjusted accordingly---a single 60 mm sabot was used for the 20 m/s case, while two 20 mm sabots were used for the 35 m/s case. To ensure repeatability, each experiment was performed three times under identical conditions, using a new dart for every trial to maintain consistent tip geometry.

The impact configuration is illustrated in Fig.~\ref{fig:exp_config}(b). To contain fragments, the back and front faces of the glass were covered with adhesive tape and paper, respectively. The sample was mounted on a wooden support within the target chamber, aligned for central impact. A stopper at the barrel exit arrested the sabot and the thicker section of the projectile, allowing only the thin cylindrical dart (up to 20 mm) to penetrate the target.

Post-experiment, the fragmented sample (Fig.~\ref{fig:exp_config}(b), right panel) was collected and imaged with uniform backlighting. The obtained images were found to have a resolution of 1 pixel $\approx 0.018$ mm. Standard image processing techniques—background removal, masking, enhancement, and edge detection—were applied to identify and characterize the resulting fragments, as shown in the sequence in Fig.~\ref{fig:exp_config}(c). This allowed us to obtain detailed fragment statistics as described in the results section.

\subsection{Development of residually-stressed network model}\label{sec:glass_model}

We approximate the three-dimensional (3D) fragmentation problem as a two-dimensional (2D) mid-plane fracture process. This simplification is appropriate for glass thicknesses $\leq 5$ mm, as we establish \emph{a posteriori}, see Sec.~\ref{sec:discussion}.

\subsubsection{Eigenstrains and residual stress formulation}
We model residual stress by introducing an incompatible inelastic strain field, $\boldsymbol{\epsilon}^*$ \cite{mura2013micromechanics}. This inelastic strain induces an associated elastic strain, $\boldsymbol{\epsilon}^e$, such that the total compatible strain $\boldsymbol{\epsilon}$ is:
\begin{align}
	\boldsymbol{\epsilon} = \boldsymbol{\epsilon}^e + \boldsymbol{\epsilon}^*
\end{align}
Where the total strain is assumed linear in the displacement gradient as usual. The residual stress $\boldsymbol{\sigma}^e$ is given by:
\begin{align}
	\boldsymbol{\sigma}^e = \mathbf{C}:(\boldsymbol{\epsilon} - \boldsymbol{\epsilon}^*) = \boldsymbol{\sigma} - \mathbf{T}^*\label{eq:residualstress}
\end{align}
where $\boldsymbol{\sigma} = \mathbf{C}:\boldsymbol{\epsilon}$ is the total stress and $\mathbf{T}^* = \mathbf{C}:\boldsymbol{\epsilon}^*$ is the apparent inelastic stress. For an isotropic solid, $C_{ijkl} = \lambda \delta_{ij}\delta_{kl} + 2\mu \delta_{ik}\delta_{jl}$ and under plane strain conditions, $\lambda/\rho = c_l^2 - 2c_s^2$ and $\mu/\rho = c_s^2$, where $c_l$ and $c_s$ are the longitudinal and shear wave speeds, dependent only on Young's modulus $E$, density $\rho$, and Poisson's ratio $\nu$ \cite{achenbach2012wave}.

The residual stress field (Eq.~\ref{eq:residualstress}) must satisfy equilibrium and traction-free boundary conditions so that
\begin{align}
	\boldsymbol{\nabla}\cdot\boldsymbol{\sigma} = 	\boldsymbol{\nabla}\cdot\mathbf{T}^* \hspace{0.5cm}\text{and}\hspace{0.5cm} \boldsymbol{\sigma} \mathbf{\hat{n}} = \mathbf{T}^* \mathbf{\hat{n}} \label{eq:resi_equil_eq}
\end{align}
This reduces the problem to solving for total stress $\boldsymbol{\sigma}$ subject to a body force $\boldsymbol{\nabla}\cdot\mathbf{T}^*$ and surface traction $\mathbf{T}^*\mathbf{\hat{n}}$. Expressed in terms of displacement, the governing Navier-Lam\'{e} equation is:
\begin{align}
	(c_l^2 - c_s^2)\boldsymbol{\nabla}(\boldsymbol{\nabla}\cdot\mathbf{u}) + c_s^2 \boldsymbol{\nabla}^2 \mathbf{u} - \frac{1}{\rho} \boldsymbol{\nabla}\cdot \mathbf{T}^* = 0 \label{eq:navier}
\end{align}

\begin{figure}[h!]
	\centering
	\includegraphics[width = 0.9\columnwidth]{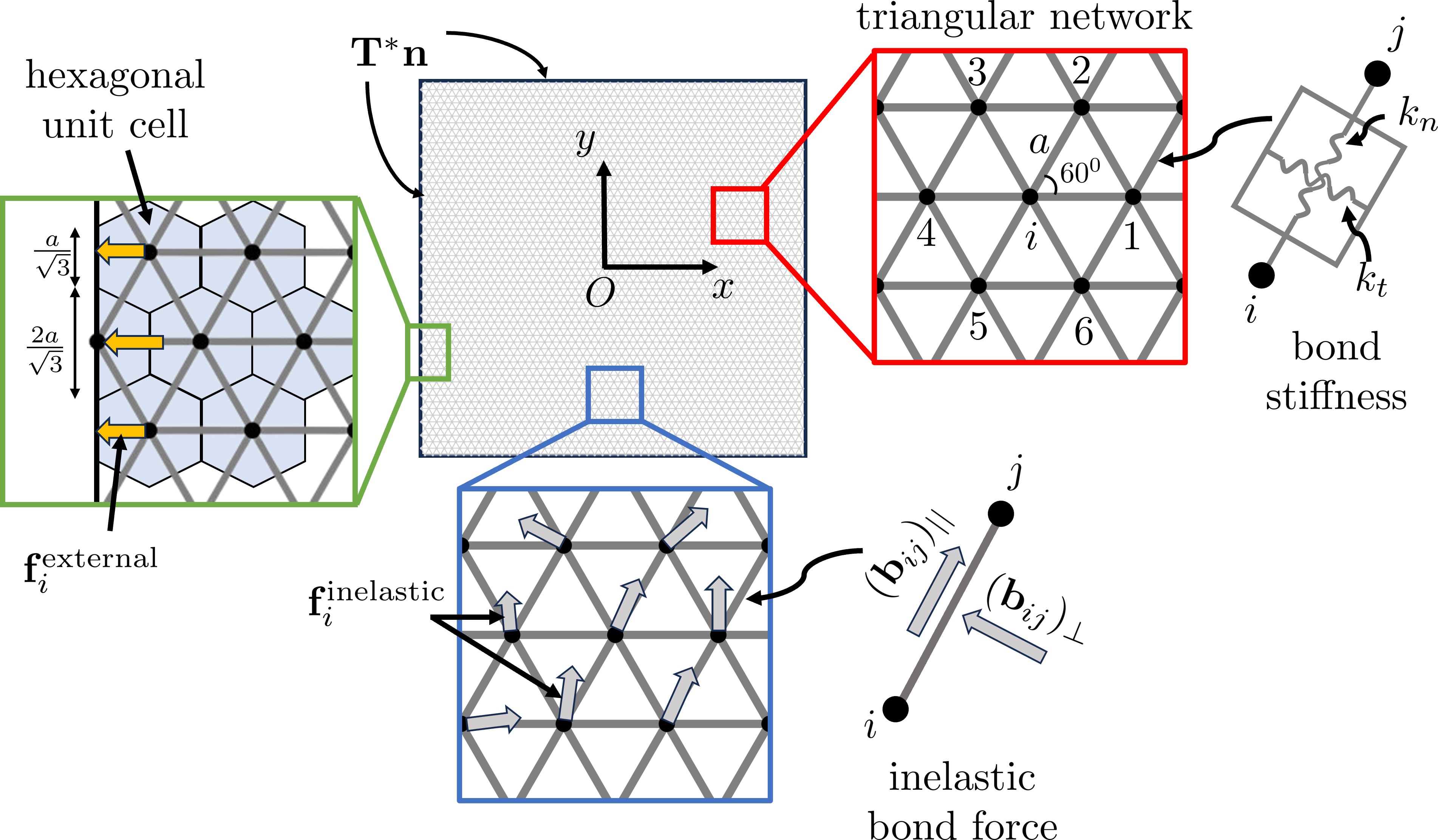}
	\caption{Schematic of the triangular network discretization. Insets depict: (Right-Red) network connectivity and bond stiffness; (Bottom-Blue) inelastic nodal forces; and (Left-Green) surface forces on boundary nodes.}
	\label{fig:network_sch}
\end{figure}

The 2D domain is discretized in the form of a triangular lattice, see Fig.~\ref{fig:network_sch}. The finite difference form of Eq.~\ref{eq:navier} at node $i$ is:
\begin{align}
	\sum_{j=1}^6\left[\frac{4(c_l^2 - c_s^2/3)}{3a^2}(\mathbf{u}_{ij}\cdot \mathbf{\hat{r}}_{ij}) \mathbf{\hat{r}}_{ij} + \frac{c_s^2 - c_l^2/3}{a^2}\mathbf{u}_{ij} \right] - \frac{1}{3a\rho}\sum_{j}^6 \mathbf{T}_{ij}^* \mathbf{\hat{r}}_{ij} = 0
\end{align}
Here, the summation is over the six nearest neighbors $j = 1\ldots 6$, $\mathbf{u}_{ij} = \mathbf{u}_j - \mathbf{u}_i$ is the relative displacement, and $\mathbf{\hat{r}}_{ij} = (\mathbf{r}_j - \mathbf{r}_i)/|\mathbf{r}_j - \mathbf{r}_i|$ is the unit bond vector. $\mathbf{T}_{ij}^* = \mathbf{T}_j^* - \mathbf{T}_{i}^*$ represents the difference in apparent inelastic stress. In terms of spring constants between adjacent nodes, this can be written as:
\begin{align}
	\sum_{j}^6\left[(k_n-k_t)(\mathbf{u}_{ij}\cdot \mathbf{\hat{r}}_{ij}) \mathbf{\hat{r}}_{ij} + k_t \mathbf{u}_{ij}  -  \mathbf{b}_{ij} \right] = 0 \label{eq:modeleq}
\end{align}
where the normal ($k_n$) and transverse ($k_t$) stiffnesses are:
\begin{align}
	k_n = \frac{c_l^2 - c_s^2/3}{a^2},\hspace{0.5cm} k_t = \frac{c_s^2 - c_l^2/3}{a^2} \label{eq:stiffness}
\end{align}
and the inelastic bond force is:
\begin{align}
	\mathbf{b}_{ij} = \frac{1}{3a\rho}\mathbf{T}_{ij}^*\mathbf{\hat{r}}_{ij} \label{eq:inelastic_bond_force}
\end{align}
The total force at node $i$ decomposes into the elastic bond force $\mathbf{f}_i^{\text{bond}}$ and the inelastic force $\mathbf{f}_i^{\text{inelastic}}$:
\begin{align}
	\mathbf{f}_i^{\text{bond}} &= \sum_{j}^6\left[(k_n-k_t)(\mathbf{u}_{ij}\cdot \mathbf{\hat{r}}_{ij}) \mathbf{\hat{r}}_{ij} + k_t \mathbf{u}_{ij} \right] \label{eq:bond_force} \\
	\mathbf{f}_i^{\text{inelastic}} &= \sum_{j}^6 \mathbf{b}_{ij} = \sum_{j}^6 \frac{1}{3a\rho}\mathbf{T}_{ij}^*\mathbf{\hat{r}}_{ij} \label{eq:inelastic_force}
\end{align}
Thus, residual stresses are incorporated by applying nodal forces $\mathbf{f}_{i}^\text{inelastic}$ corresponding to the prescribed $\boldsymbol{\epsilon}^*$, see blue inset in Fig.~\ref{fig:network_sch}.

\begin{algorithm}[H]
\caption{Residual stress simulation procedure}\label{alg:simulations}
\begin{algorithmic}[1]

\Statex \textbf{Part I: Static Simulation}
\State Initialize network topology
\State Assign stiffness parameters $(k_n, k_t)$ \Comment{Using Eq.~\ref{eq:stiffness}}
\State Apply inelastic forces $\mathbf{f}_i^{\mathrm{inelastic}}$ \Comment{Using Eq.~\ref{eq:inelastic_force}}
\State Apply external forces $\mathbf{f}_i^{\mathrm{external}}$ \Comment{$\mathbf{T}^*\mathbf{\hat{n}}$ at the boundaries}
\State \Call{Equilibrate}{Network} \Comment{Solve using Conjugate Gradient Method}
\State \textbf{Output:} Residual-stressed network
\Statex
\Statex \textbf{Part II: Dynamic Simulation}
\State Import residual-stressed network
\State Cut-off central hole \Comment{Modify topology}
\State Initialize time $t \gets 0$
\State $\text{topology\_changed} \gets \text{True}$ \Comment{Force initial update}

\Loop
    \If{$\text{topology\_changed}$}
        \State Update $\mathbf{f}_i^{\mathrm{inelastic}}$ \Comment{via Eq.~\ref{eq:inelastic_force} for intact bonds}
        \State $\text{topology\_changed} \gets \text{False}$
    \EndIf

    \State Update $\mathbf{f}_i^{\mathrm{bond}}$ \Comment{Using Eq.~\ref{eq:bond_force}}
    \State Apply $\mathbf{u}_i(t)$ to hole boundary nodes
    \State Obtain $\mathbf{u}_i(t + \Delta t)$ \Comment{Time integration step, Eq.~\ref{eq:model_dyneq}}

    \State Compute maximum principal bond strain $\epsilon^{\text{max}}_{ij}$ \Comment{Using Eq.~\ref{eq:nodalstrain}}
    \If{$\epsilon^{\text{max}}_{ij} \geq \epsilon_b$}   \Comment{Check critical strain threshold}
        \State Set $k_n \gets 0$, $k_t \gets 0$ \Comment{Break bond}
        \State $\text{topology\_changed} \gets \text{True}$ \Comment{Trigger inelastic update}
    \EndIf
    \If{$t > t_{\max}$}
        \State \textbf{Stop}
    \EndIf
    \State $t \gets t + \Delta t$
\EndLoop
\end{algorithmic}
\end{algorithm}


\subsubsection{Dynamic formulation and benchmarking}
The dynamic evolution of the system is governed by the equilibrium equation with an added inertial term, subject to traction-free boundary conditions:
\begin{align}
	\rho\frac{\partial^2 \mathbf{u}}{\partial t^2} = \boldsymbol{\nabla}\cdot\boldsymbol{\sigma} - \boldsymbol{\nabla}\cdot\mathbf{T}^* \hspace{0.5cm}\text{and}\hspace{0.5cm} \boldsymbol{\sigma}\mathbf{\hat{n}} = \mathbf{T}^*\mathbf{\hat{n}} \label{eq:resi_dyn_eq}
\end{align}
Discretization yields the equation of motion for node $i$:
\begin{align}
	\frac{\partial^2 \mathbf{u}_{i}}{\partial t^2} = \mathbf{f}_i^\text{bond} - \mathbf{f}_i^\text{inelastic} \label{eq:model_dyneq}
\end{align}
The total force on node $i$ includes external forces and damping:
\begin{align}
	\mathbf{f}_{i} = \mathbf{f}_i^{\text{bond}} - \mathbf{f}_i^{\text{inelastic}} + \mathbf{f}_i^{\text{external}} - \zeta\frac{\partial \mathbf{u}_i}{\partial t}
\end{align}
where $\mathbf{f}_i^{\text{external}}$ accounts for surface tractions $\mathbf{T}^*\mathbf{\hat{n}}$ at boundaries, and $\zeta$ is a numerical damping constant. The system is evolved using the Verlet algorithm:
\begin{align}
	\mathbf{u}_{i}(t+\Delta t) = 2\mathbf{u}_{i}(t) - \mathbf{u}_{i}(t-\Delta t) + \mathbf{f}_{i}\Delta t^2 \label{eq:verlet}
\end{align}
For stability, we use $\zeta = 0.1$ and $\Delta t = 0.45a/c_r$ \cite{del2002wave,braun2014new}.

To ensure the validity of this scheme, it was extensively benchmarked against against both standard elastodynamic solutions and LEFM results, as described in \suppMat\suppref{S1}.

\subsubsection{Simulation Parameters and Protocol}
\label{subsec:simulationParams}

\begin{figure}[h!]
	\centering
	\includegraphics[width = \columnwidth]{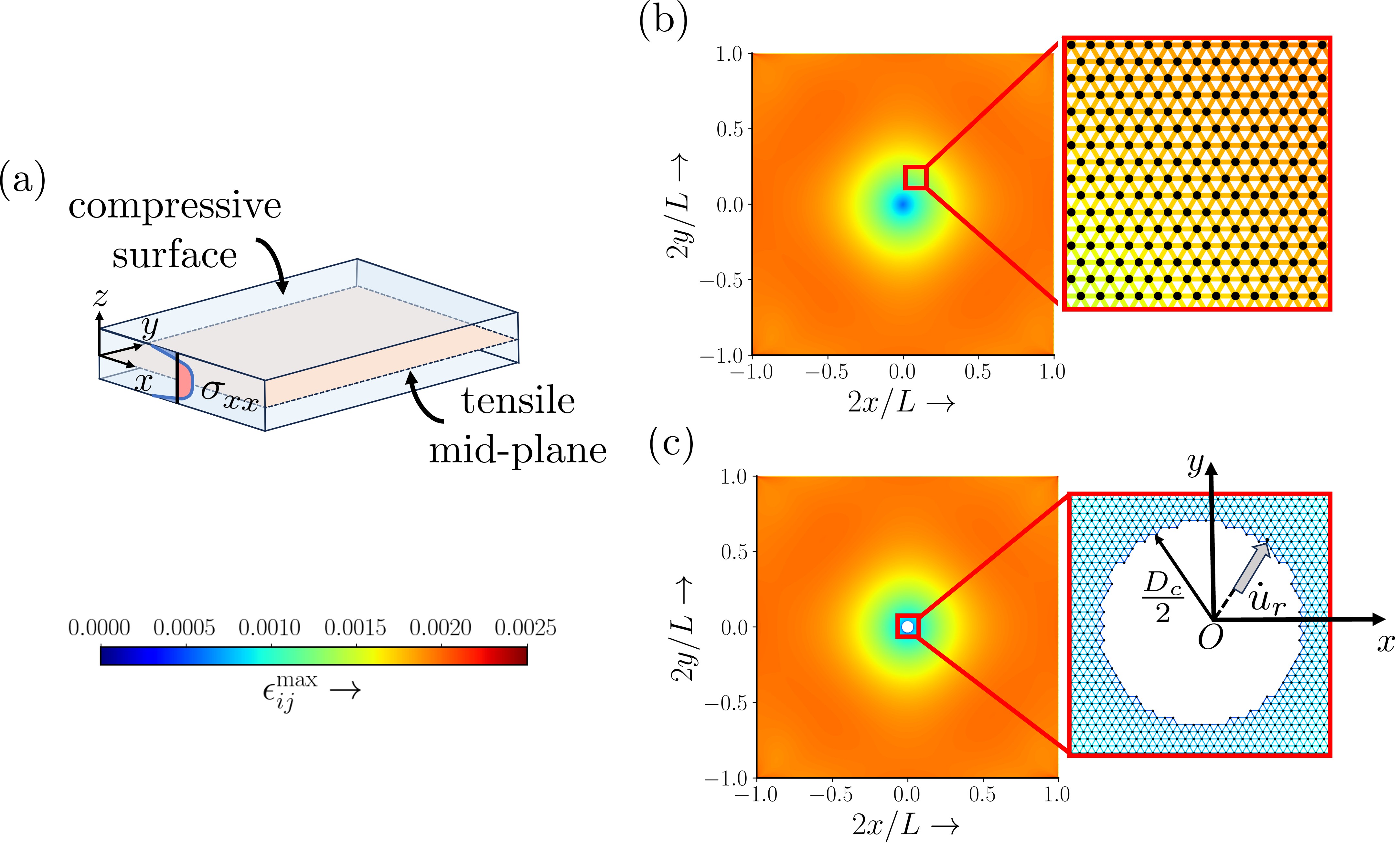}
	\caption{(a) Schematic showing the simulated tensile mid-plane, (b) Residual stress field after static equilibration. (c) Crack initiation via hole excision and radial expansion.}
	\label{fig:dynamic_steps}
\end{figure}

The simulation framework was applied to the mid-plane of the residually stressed glass sample, see schematic in Fig.~\ref{fig:dynamic_steps}(a). While the thickness of the plate shows the typical parabolic profile, we focus on the tensile mid-plane, which is responsible for generating the quasi-2D fracture pattern, \emph{cf.} Fig.~\ref{fig:exp_config}(c)

A square domain ($L = 60$ mm) is discretized into a triangular network of $601 \times 694$ nodes with lattice spacing $a = 0.1$ mm. Distance and velocity are normalized by $L/2$ and the Rayleigh wave speed $c_r$, respectively. Normalization renders the results dependent only on Poisson's ratio (set to $\nu = 0.2$ for glass).

We assume an axisymmetric inelastic strain distribution $\boldsymbol{\epsilon}^* (r)$ relative to the center:
\begin{align}
	\epsilon_{rr}^*(r) = \epsilon_{\theta\theta}^*(r) = \begin{cases}
		\epsilon_0^* \left[1-\left(1-\frac{r}{R}\right)^m\right] & r/R\leq 1\\
		\epsilon_0^* & r/R > 1
	\end{cases} \label{eq:inelastic_strain}
\end{align}
with $\epsilon_{r\theta}^* = 0$. Here, $R = 0.84L$ is the characteristic variation length, $\epsilon_0^*$ is the maximum strain, and $m$ controls the profile shape.

For each bond, we define a strain-based fracture criterion as follows. First, the total strain tensor $\boldsymbol{\epsilon}_{i}$, at node $i$, is:
\begin{align}
	\boldsymbol{\epsilon}_{i} = \frac{1}{6a}\sum_{j}\left[\mathbf{u}_{ij}\otimes \mathbf{\hat{r}}_{ij} + \mathbf{\hat{r}}_{ij}\otimes \mathbf{u}_{ij}\right] \label{eq:nodalstrain}
\end{align}
The strain of the bond connecting nodes $i$ and $j$ is interpolated as the average of $\boldsymbol{\epsilon}_i$ and $\boldsymbol{\epsilon}_j$. A bond breaks irreversibly if its maximum principal strain, $\epsilon_{ij}^\text{max}$, exceeds a threshold $\epsilon_b$. Since this criterion depends on the macroscopic strain interpolated from the nodal strain field, crack propagation is driven by mechanics. As will be discussed in Sec.~\ref{subsec:simulated_fracture}, this criterion ensures that macroscopic crack propagation remains independent of mesh topology and lattice direction \cite{martin2005mechanisms, levandovsky2007mechanisms}.

Simulations proceed in two stages, see algorithm~\ref {alg:simulations}. First, a static simulation, with no bond-breaking, is used to generate the residual stress field. Inelastic nodal forces (Eqs.~\ref{eq:inelastic_bond_force}-\ref{eq:inelastic_force}) and boundary surface forces (derived from traction $\mathbf{T}^*\mathbf{\hat{n}}$ integrated over boundary node Voronoi cells) are applied. The network is relaxed to equilibrium using the Conjugate Gradient Method. The resulting pre-stressed network is shown in Fig.~\ref{fig:dynamic_steps}(b).

Following this, a dynamic simulation is used to study fracture arising from microscopic bond breaking. In order to simulate the projectile impact, a circular hole of diameter $D_c = 2.4$ mm is first introduced at the center. The nodes on the hole boundary are displaced radially outward at a rate $\dot{u}_r$ (Fig.~\ref{fig:dynamic_steps}(c)), simulating the penetration of the $43^\circ$ projectile tip, \emph{cf.} Fig.~\ref{fig:exp_config}. Bond breaking is enabled, and Eq.~(\ref{eq:verlet}) is solved with simultaneous breaking criterion evaluation. 

Table~\ref{Tab:sim_parameter} summarizes the simulation cases. In cases 1 and 2, the loading rate $\dot{u}_r$ is varied, corresponding to impact speeds of 20 and 35 m/s, in cases 2–4, the inelastic strain amplitude $\epsilon_0^*$ is changed while in cases 2, 5, and 6 have varying strain profile parameter $m$.

\begin{table}[h!]
	\renewcommand{\arraystretch}{1.2}
	\centering
	\caption{Simulation parameters for the studied configurations.}
	\label{Tab:sim_parameter}
	\begin{tabular}{c c c c c c} 
		\hline
		\textbf{Case} & $\boldsymbol{\dot{u}_r}$ (m/s) & $\boldsymbol{m}$ & $\boldsymbol{\epsilon_b/\epsilon_0^*}$  & $\boldsymbol{\epsilon_0^*}$ &$\boldsymbol{\sigma}^e_\text{max}$ (MPa)   \\
		\hline
		1 & 8 & 5 & 1.5 & 0.002 & 55.6   \\
		2 & 15 & 5 & 1.5 & 0.002 & 55.6   \\
		3 & 15 & 5 & 3 & 0.001  & 27.8 \\
		4 & 15 & 5 & 6 & 0.0005  & 14.0 \\
		5 & 15 & 3 & 1.5 & 0.002 & 48.3   \\
		6 & 15 & 2 & 1.5 & 0.002 & 40.5 \\
		\hline
	\end{tabular}
\end{table}

\section{Results}
\label{sec:results}

In this section, we analyze the fracture and fragmentation characteristics obtained from both projectile impact experiments and network-based simulations. We first characterize the experimental fragment size distributions and fracture patterns, then present the corresponding numerical results. Finally, we investigate the underlying mechanisms of fragment formation, exploring the influence of residual stress parameters and microscopic crack dynamics—specifically speed and branching—within the network framework.

\subsection{Experimental glass fragments and fracture pattern}
\begin{figure}[h!]
	\centering
	\includegraphics[width = \columnwidth]{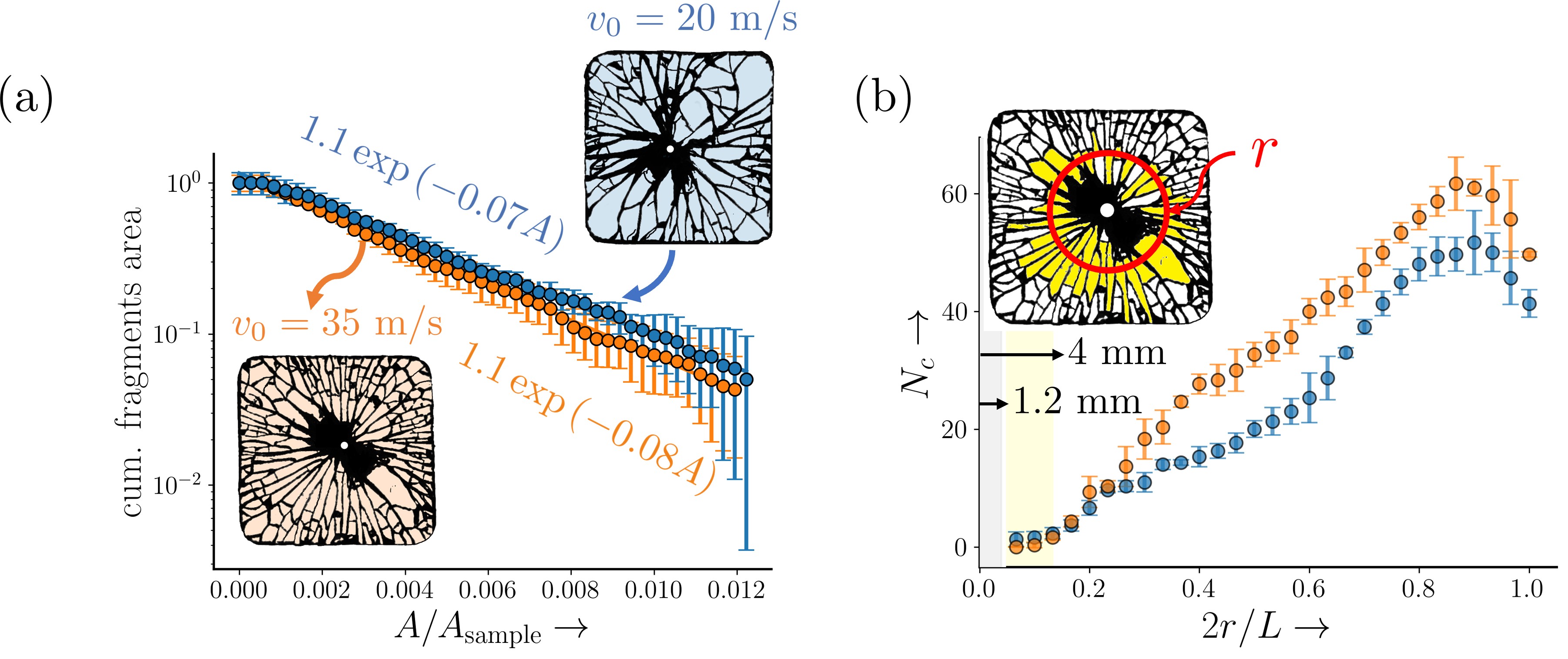}
	\caption{(a) Cumulative fragment area distribution with insets showing representative fracture patterns. (b) Radial variation of fragment count, $N_c(r)$. Blue and orange data correspond to impact velocities of 20 and 35 m/s, respectively. The inset in (b) shows the fragments (solid yellow) that contribute to $N_c(r)$ at a given $r$ (red circle).}
	\label{fig:exp_frag_size}
\end{figure}

The macroscopic response of the toughened glass to projectile impact is visualized in Figure \ref{fig:exp_frag_size}. Panel (a) contrasts the fracture topologies obtained at two distinct impact velocities: $v_0 = 20$ m/s (depicted in blue) and $v_0 = 35$ m/s (in orange), with the central white circle demarcating the impact zone of the 2.4 mm dart. To isolate the contribution of stored elastic energy, we refer to control experiments on non-tempered glass (\suppMat\suppref{S2}), which confirm that the dense fragmentation observed here is governed primarily by the release of residual stress rather than local contact mechanics. A direct comparison of the insets reveals a clear morphological transition: the fracture network shifts from a coarser structure at lower velocities to a significantly finer mesh at higher velocities. Although such variations are sometimes ascribed to spatial fluctuations in the residual stress profile \cite{tandon2015fracture, zaccaria2020nondestructive}, the reproducibility of this trend across three independent trials provides strong evidence that the observed refinement is a deterministic function of the impact velocity.

Quantitatively, the cumulative fragment area distribution—defined as the fraction of fragments with an area greater than or equal to $A$—follows an exponential decay for both impact speeds (Fig. \ref{fig:exp_frag_size}(a)). Fragments with $A < 2$ mm$^2$ ($\approx 3\%$ of the maximum fragment area) were excluded to eliminate potential artifacts from lighting noise, despite the high image resolution (1 pixel $\approx 0.018$ mm). Remarkably, despite the visual difference in pattern coarseness, the exponential fitting parameters for both speeds are nearly identical. The solid lines in Fig.~\ref{fig:exp_frag_size}(a) represent the exponential fit to the mean cumulative area from three trials, with error bars indicating the standard deviation.

To reconcile the visual difference in fracture density with the similarity in global size distribution, we analyzed the spatial topology of the cracks. This was done by computing $N_c(r)$, the number of fragments intersecting a circle of radius $r$ centered at the impact point, as depicted in Fig. \ref{fig:exp_frag_size}(b) (inset). This panel plots the variation of $N_c(r)$ with $r$ where markers represent the mean of three trials and error bars denote the standard deviation. Data within the projectile radius ($r \leq 1.2$ mm) and the imaging-limited zone ($r \leq 4$ mm, or $2r/L < 0.2$) are shaded gray and black, respectively. Beyond this near-field region ($2r/L > 0.2$), the higher impact speed (35 m/s) clearly yields a higher fragment count compared to the lower speed (20 m/s).

These results highlight a nuance in characterizing fragmentation: while standard metrics like global size distribution—often measured far from the impact center \cite{pourmoghaddam2018experimental}—suggest invariance with respect to velocity, the radial fragment count $N_c(r)$ captures significant topological differences in the fracture network.

\subsection{Simulated fracture pattern and fragment-size distribution}\label{subsec:simulated_fracture}

\begin{figure}[H]
	\centering
	\includegraphics[width = \columnwidth]{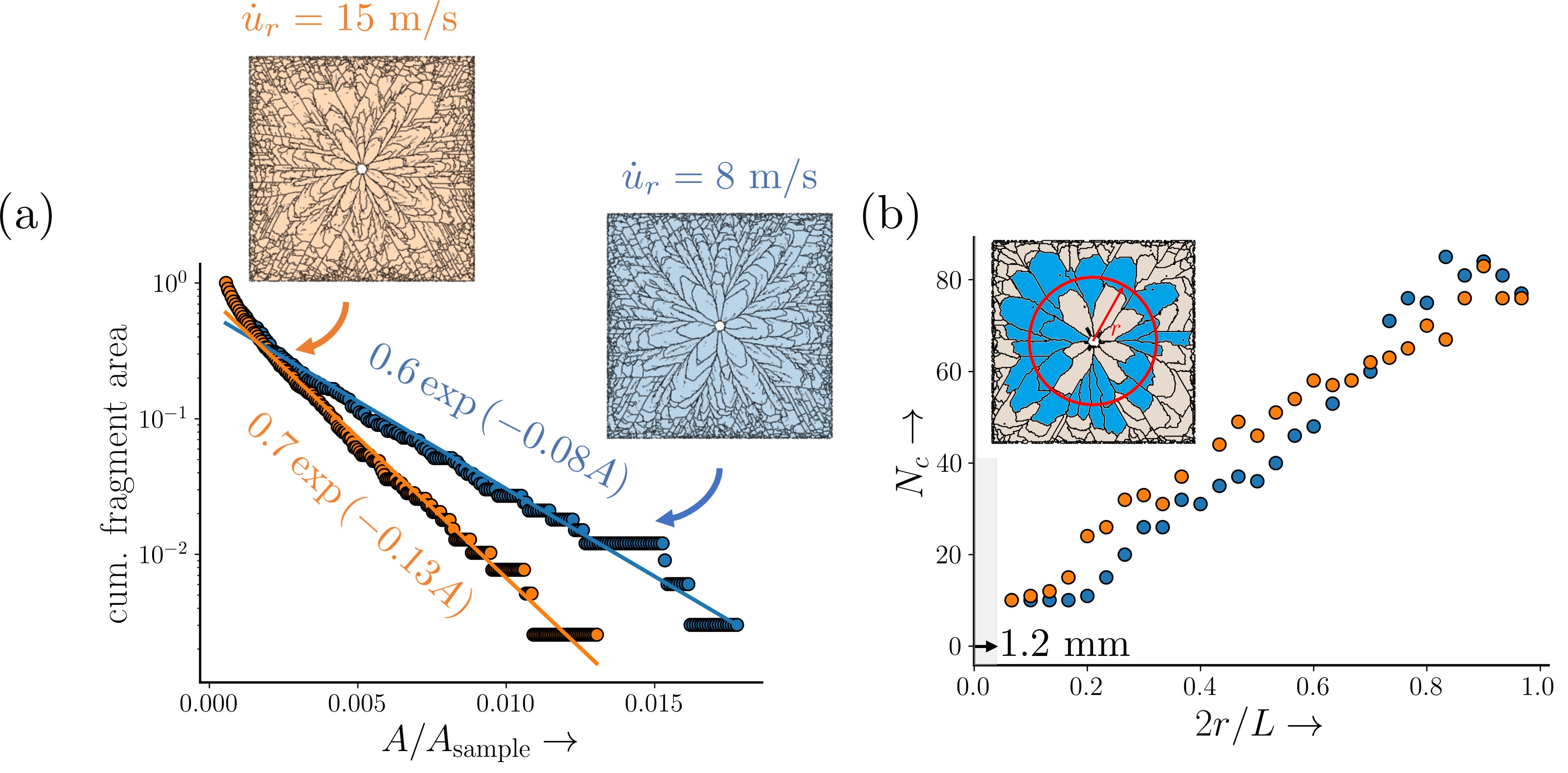}
	\caption{(a) Simulated cumulative fragment area distribution with insets showing representative fracture patterns. (b) Radial variation of fragment counts, $N_c(r)$. Blue and orange data correspond to radial loading rates of 8 m/s and 15 m/s, respectively (Cases 1 and 2 in Table~\ref{Tab:sim_parameter}). The inset in (b) shows the fragments (solid blue) that contribute to $N_c(r)$ at a given $r$ (red circle).}
	\label{fig:num_frag_size}
\end{figure}

The experimentally observed influence of the impact velocity on the resulting fracture topology is also captured by complementary numerical simulations, see Fig.~\ref{fig:num_frag_size}. Panel (a) specifically compares the patterns generated under two distinct radial displacement rates, $\dot{u}_r = 8$ m/s (blue) and $15$ m/s (orange), corresponding to Cases 1 and 2 in Table~\ref{Tab:sim_parameter}. For these simulations, we utilized an inelastic strain profile with shape parameter $m = 5$ (\emph{cf.} Sec.~\ref{subsec:simulationParams}) and established a breaking threshold of $\epsilon_b = 0.003$. This threshold was selected to be marginally higher than the maximum pre-existing strain ($\epsilon_{ij}^\text{max} = 0.0026$) within the static network, ensuring stability prior to dynamic loading. A closer examination of the insets in Fig.~\ref{fig:num_frag_size}(a) reveals a clear rate-dependency: the higher loading rate induces a greater density of primary fractures around the central impact zone, producing 13 elongated fragments compared to 10 for the lower rate. Spatially, both cases exhibit a characteristic cascading size---transitioning from coarse fragments near the impact center to finer fragmentation towards the free boundaries.

Quantitatively, the cumulative fragment area distribution follows an exponential decay ($\approx ae^{-bA}$) for both loading rates (Fig. \ref{fig:num_frag_size}(a)), mirroring the experimental trends in Fig. \ref{fig:exp_frag_size}(a). However, unlike the experiments, the decay parameter $b$ in the simulations shows a clear dependence on loading rate. It is also observed that the simulations generate a significant population of fine fragments ($A < 2$ mm$^2$) near the boundaries. To maintain consistency with the experimental resolution limits, these fine fragments are excluded from the statistical analysis presented here.

Analogous to Fig.~\ref{fig:exp_frag_size}, the radial variation of fragment count, $N_c(r)$ for the numerical simulation is shown in Fig.~\ref{fig:num_frag_size}(b). A key advantage of the numerical model is its ability to resolve the nucleation region near the impact site, a zone often obscured in experiments. The model captures a higher number of nucleating cracks for the higher loading rate (15 m/s) compared to the lower rate (8 m/s). As the cracks propagate outward ($2r/L \approx 0.8$), the distinction between the two rates diminishes, and the $N_c$ values become indistinguishable.

The simulated fracture patterns show strong qualitative agreement with experimental observations, particularly in the formation of \lq flower-like\rq\ fragments near the load point. This morphology is consistent with both our experimental data (see \suppMat\suppref{S3}) and high-fidelity 3D simulations reported in the literature \cite{stewart2023modeling}. Furthermore, the model captures the reduction in fragment size near the boundaries, a phenomenon attributed to stress wave reflections from the free edges \cite{zaccaria2020nondestructive, zijlstra1969fracture, tandon2015fracture}.

Finally, regarding the numerical discretization, it is important to note that while the triangular lattice geometry restricts local bond breaking to six specific directions, the macroscopic crack paths are not artificially aligned with the mesh. Because the failure criterion depends on the macroscopic strain tensor—interpolated from the nodal field—the crack trajectories emerge naturally from the mechanics rather than the grid topology. Consequently, we expect that using an alternative mesh geometry (e.g., square) would yield statistically similar macroscopic fracture patterns. 

\subsection{How does crack branching lead to fragment formation?}\label{subsec:fragmentform}

\begin{figure}[h!]
	\centering
	\includegraphics[width = \columnwidth]{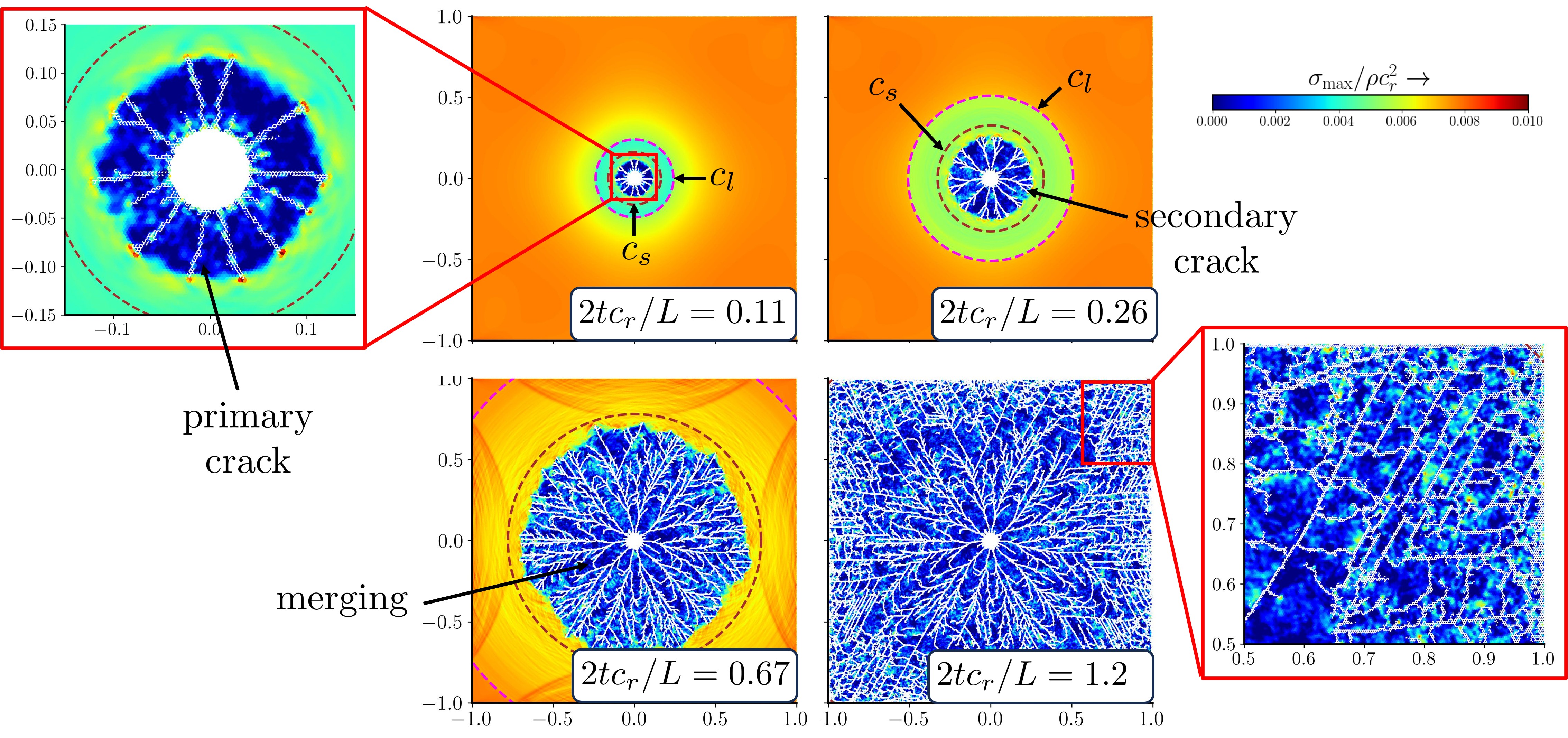}
	\caption{Simulated fracture evolution at four time snapshots for Case 2 in Table~\ref{Tab:sim_parameter}.}
	\label{fig:field_snaps}
\end{figure}

The temporal evolution of the fracture process is detailed in Figure \ref{fig:field_snaps}, which presents a sequence of snapshots for the representative Case 2 (see Table~\ref{Tab:sim_parameter}). To visualize the driving mechanical forces, the developing fracture network in these frames is overlaid with the maximum principal stress field, $\sigma_{\text{max}}$, normalized by $\rho c_r^2$.

During the initial phase ($2t c_r/L \leq 0.11$), the radial displacement loading triggers multiple bond-breaking events near the hole boundary. However, as shown in the left inset, only approximately eleven primary cracks successfully nucleate and propagate radially, characterized by stress concentrations at their tips. By $2t c_r/L = 0.26$, these primary cracks bifurcate into multiple secondary branches. Macroscopically, the crack tip velocities remain subsonic relative to the longitudinal wave speed $c_l$, as evidenced by the clear separation between the crack tips and the expanding $c_l$ and $c_s$ wavefronts in the stress field.

As the simulation progresses to $2t c_r/L = 0.67$, secondary cracks from adjacent primary branches begin to merge, isolating fragments, while others continue to bifurcate. The branching rate notably increases as the cracks approach the domain boundaries. At this stage, longitudinal waves reflected from the free edges interact with the propagating fracture network. By $2t c_r/L = 1.2$, fragmentation is complete with no further nucleation. The complex interplay between crack-crack interactions and reflected stress waves ultimately generates a dense network of fine, intricate fragments near the boundaries (see right inset).

\subsection{Which parameters critically determine the fracture pattern?}\label{sec:exploit_network}
The model's ability to reproduce experimental fracture patterns and capture the underlying dynamics validates its use as a predictive tool. We now investigate how key physical parameters---specifically the residual stress magnitude (via $\epsilon_0^*$) and the stress profile shape (via $m$)---govern the resulting fragmentation topology.

\subsubsection{Residual stress magnitude regulates fragment size}
We examined the effect of residual stress intensity by varying the maximum inelastic strain $\epsilon_0^*$ while keeping the profile shape ($m=5$) and loading parameters constant. We compared Cases 2, 3, and 4 (Table~\ref{Tab:sim_parameter}), corresponding to $\epsilon_0^*$ of 0.002, 0.001, and 0.0005, and maximum residual stresses of 55.6, 27.8, and 14.0 MPa, respectively (assuming $E=70$ GPa, $\nu = 0.2$).

\begin{figure}[h!]
	\centering
	\includegraphics[width = \columnwidth]{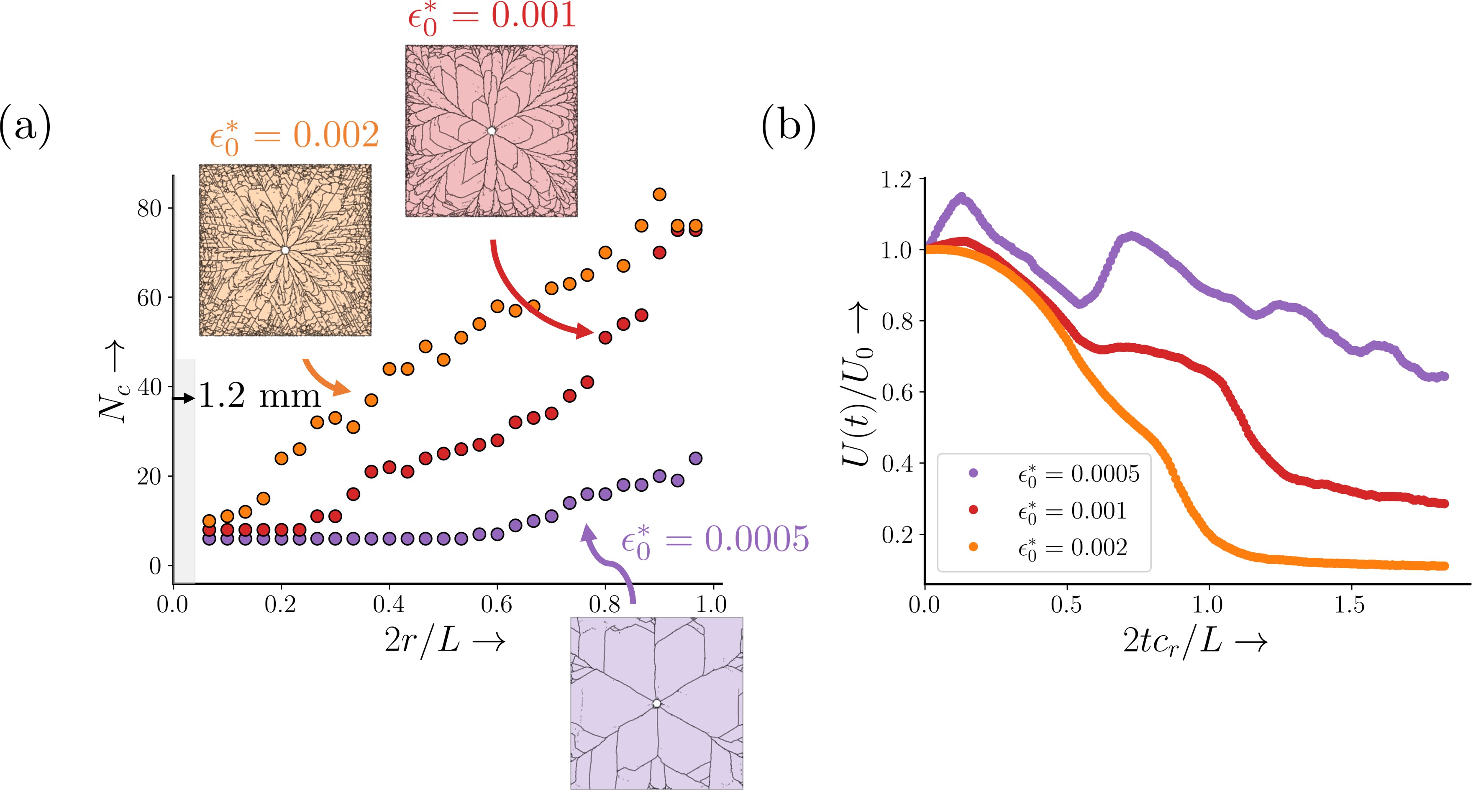}
	\caption{(a) Radial variation of fragment count $N_c(r)$ for varying inelastic strain magnitudes $\epsilon_0^*$. Insets show representative fracture patterns. Orange, red, and purple correspond to Cases 2, 3, and 4, respectively. (b) Temporal evolution of total strain energy $U(t)$ for the corresponding cases.}
	\label{fig:ep_var}
\end{figure}

The resulting fragment densities, as quantified by $N_c(r)$  for these three cases are reproduced in Fig.~\ref{fig:ep_var}(a). Insets to this panel display the corresponding patterns. As anticipated, higher residual stress (orange, $\epsilon_0^* = 0.002$) produces significantly finer fragments, while lower stress (purple, $\epsilon_0^* = 0.0005$) yields coarser fragments. The plateau in the purple curve for $r/L < 0.6$ reflects a characteristic length scale of the large fragments formed under low stress. These results align with established experimental power laws \cite{pourmoghaddam2018experimental, tandon2015fracture, zaccaria2020nondestructive} relating fragment size inversely to the magnitude of stored compressive stress \cite{barsom1968fracture, gulati1997frangibility}.

This morphological transition with increasing maximum residual strain is driven by energetics. Figure \ref{fig:ep_var}(b) plots the evolution of the total elastic strain energy $U(t)$. The system energy is comprised of two parts---energy added by external work (inelastic surface forces and boundary loading) and energy dissipated via bond breaking.
For the high residual stress cases ($\epsilon_0^* = 0.002$ and $0.001$), $U(t)$ decreases monotonically during the main fracture phase ($t < 0.5 L/c_r$), indicating that rapid, widespread fragmentation dissipates stored energy faster than it is replenished. Notably, for the highest stress case, $U(t)$ approaches zero, implying near-total release of stored elastic energy. Conversely, for the lowest stress case ($\epsilon_0^* = 0.0005$), $U(t)$ exhibits an initial rise and subsequent oscillation, indicating that the sparse fracture network is insufficient to fully dissipate the stored energy, allowing the system to retain a large part of its initial strain energy (see purple curve). Another equivalent approach to investigate a similar residual stress intensity effect is to vary the breaking criterion while keeping other parameters constant; see \suppMat\suppref{S4}.

\subsubsection{Steeper stress profiles accelerate dissipation}
\begin{figure}[h!]
	\centering
	\includegraphics[width = \columnwidth]{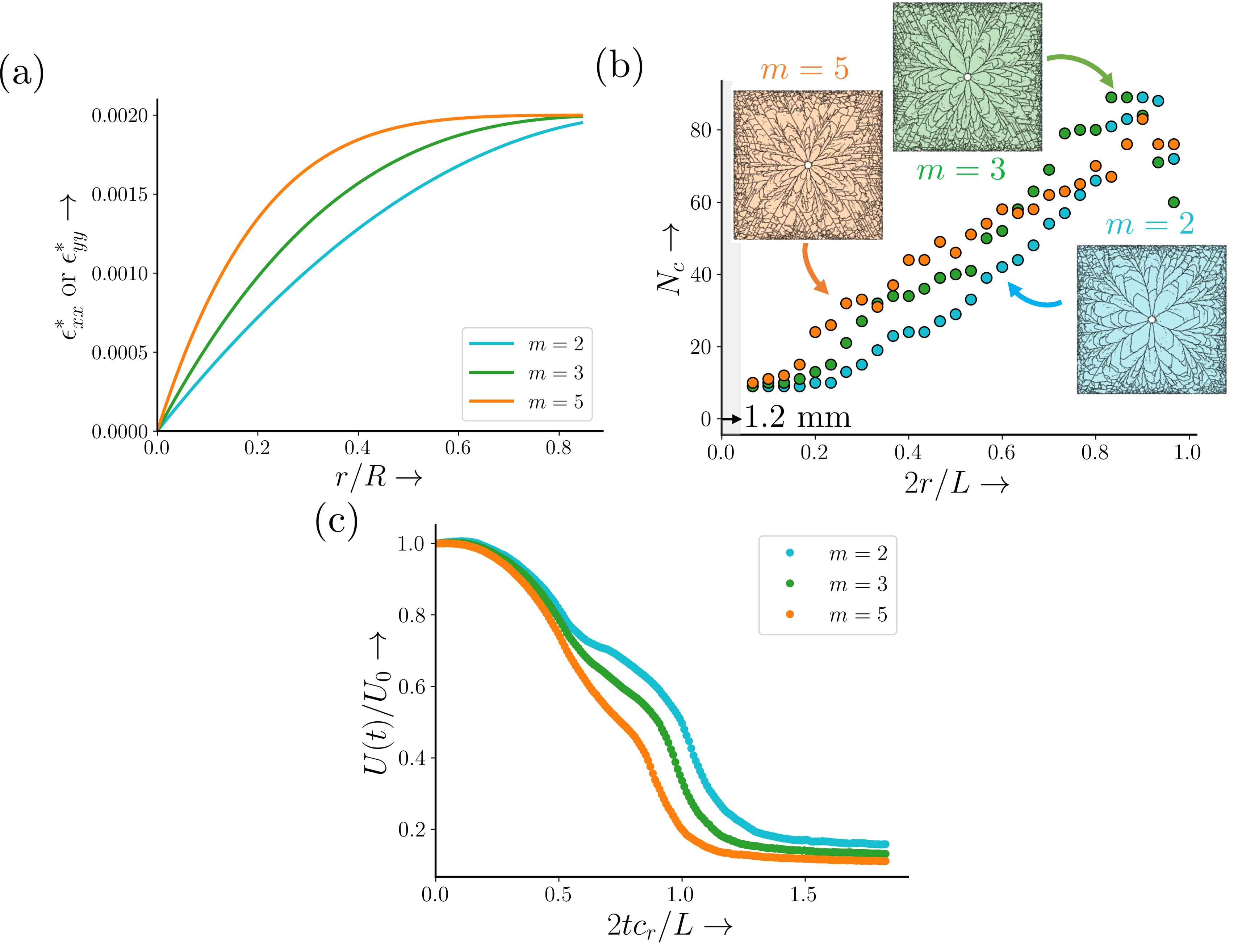}
	\caption{Effect of inelastic strain profile $\boldsymbol{\epsilon}^*(r)$. (a) Strain profiles for varying $m$. (b) Radial variation of fragment count $N_c(r)$ with fracture pattern insets. (c) Temporal evolution of total strain energy $U(t)$. Blue, orange, and green correspond to $m$ values of 2, 3, and 5, respectively. Colors in panel (b) correspond to those in panel (a).}
	\label{fig:ep_profile_var}
\end{figure}

We also analyzed the influence of the residual stress profile itself by varying the exponent $m$ (see Eq.~\ref{eq:inelastic_strain}) and keeping the breaking stress unchanged. We compared a gentle ($m = 2$, cyan), moderate ($m = 3$, green), and steep ($m = 5$, orange) gradient, corresponding to Cases 6, 5, and 2, respectively. As noted in Table~\ref{Tab:sim_parameter}, steepening the profile also results in a slight increase in the maximum induced stress (from 40.5 to 55.6 MPa). The results are summarized in Fig.~\ref{fig:ep_profile_var}; panel (a) of this figure shows the corresponding strain profiles used for this analysis. 

The fracture patterns in Fig. \ref{fig:ep_profile_var}(b) demonstrate that a steeper stress gradient produces finer fragmentation, particularly near the nucleation zone. The gentle profile ($m=2$) yields noticeably larger fragments near the impact site, whereas the steepest profile ($m=5$) generates a dense cluster of fine fragments. This is quantified by $N_c(r)$, where the steep profile sustains a higher fragment count throughout the bulk. Near the boundaries, however, edge effects dominate, rendering the distributions indistinguishable.

The corresponding system energy variation with time is reproduced in Fig.~\ref{fig:ep_profile_var}(c). The steepest profile ($m=5$) exhibits the most rapid decay in $U(t)$, confirming that a sharper stress gradient facilitates faster energy release through more intense branching.

Collectively, these results suggest two distinct pathways to control fragmentation: adjusting the magnitude of tempering (degree of ion exchange) or modifying the spatial gradient of the stress field. This implies that a glass with a lower surface compression but a steeper stress gradient—potentially achievable through tailored thermal or multi-step chemical processing—could yield finer fragmentation than a standard profile. This aligns with experimental observations where prolonged ion exchange relaxes surface stress yet alters the depth profile, significantly modifying the fracture topology \cite{tandon2015fracture, zijlstra1969fracture}.

\subsection{Microscopic crack propagation and branching}
\begin{figure}[h!]
	\centering
	\includegraphics[width = \columnwidth]{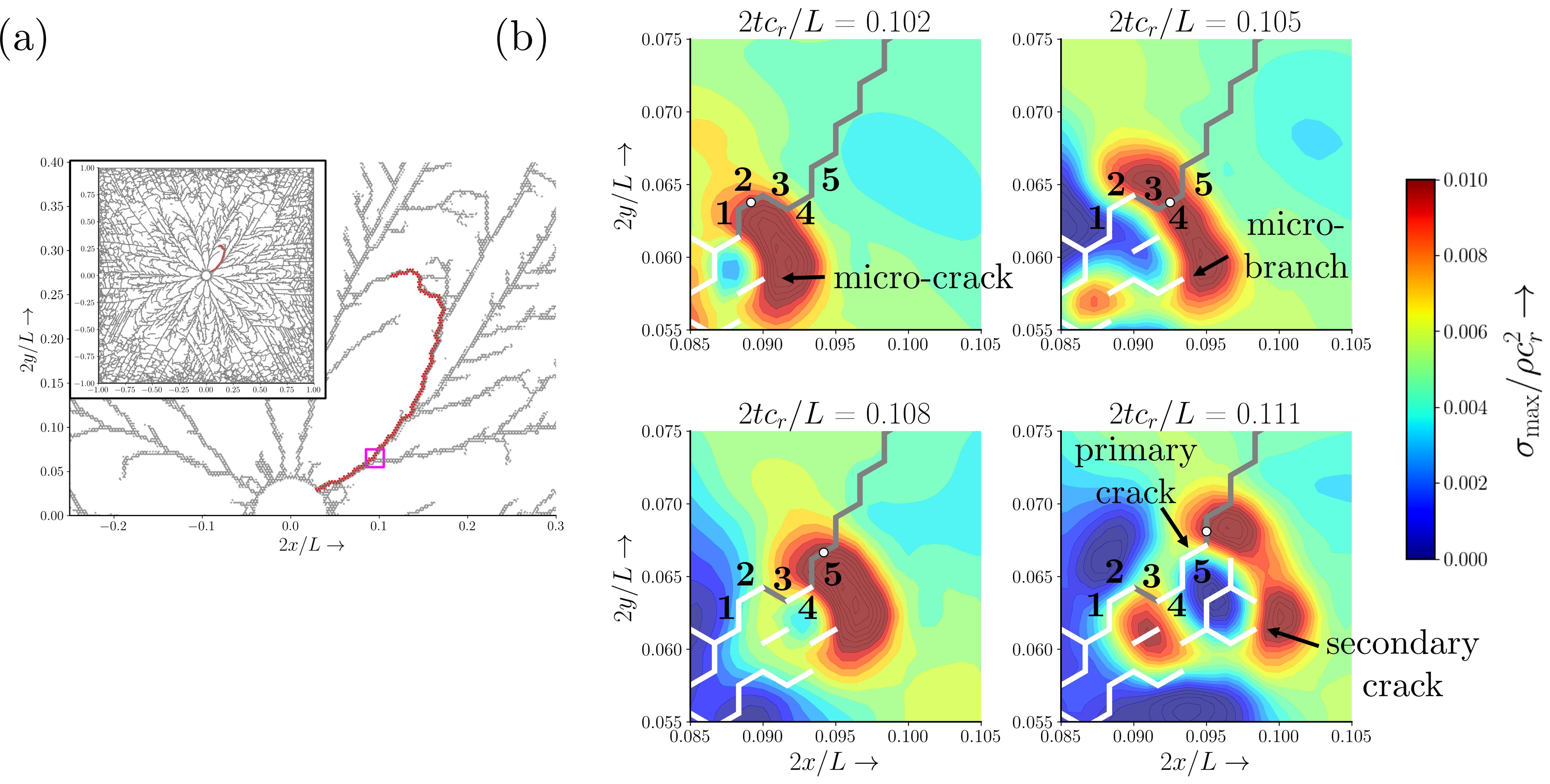}
	\caption{Microscopic bond-breaking dynamics. (a) A single crack path (red) with a highlighted segment (magenta box) showing micro-branches. (b) Temporal sequence of bond breaking events along the highlighted segment, overlaid on the maximum principal stress field.}
	\label{fig:micro_bond_break}
\end{figure}

A distinct advantage of our network-based framework is its ability to resolve the microscopic mechanisms of crack propagation, offering a window into the fundamental origins of crack branching---a sufficient, though not necessary, condition for fragmentation (as discussed in Sec.~\ref{subsec:fragmentform}). To investigate this, we isolated a single crack path from the Case 2 simulation, highlighted in red in Fig. \ref{fig:micro_bond_break}(a). As shown in the inset, numerous clusters of broken bonds (micro-cracks) form along the primary trajectory; however, most of these micro-branches remain localized and do not evolve into full secondary cracks.

To probe the dynamics at the bond level, we focus on the chain segment marked by the magenta box in the bottom right of Fig. \ref{fig:micro_bond_break}(a). Figure \ref{fig:micro_bond_break}(b) displays the sequential failure of five consecutive bonds (labeled 1–5) overlaid on the maximum principal stress field $\sigma_{\text{max}}$. White segments indicate bonds broken prior to the current snapshot, gray segments represent intact bonds, and the white marker identifies the specific bond breaking at that instant.

At $2t c_r/L = 0.102$, the bond preceding segment 1 is broken. However, the next bond to fail is not segment 1, but segment 2 (indicated by the marker). By the next snapshot ($2t c_r/L = 0.105$), both bonds 1 and 2 have failed, and bond 4 breaks ahead of bond 3. Simultaneously, bonds in the immediate neighborhood fail, initiating micro-branches. By $2t c_r/L = 0.108$, bond 5 breaks while bond 3 remains intact. This sequence demonstrates that at the microscopic scale, bond breaking is not strictly sequential; \lq daughter\rq\ cracks often nucleate ahead of the main tip and coalesce retroactively. By the end of this sequence, two distinct stress concentration zones emerge, ultimately driving the bifurcation of the primary crack into a secondary branch.

\begin{figure}[h!]
	\centering
	\includegraphics[width = \columnwidth]{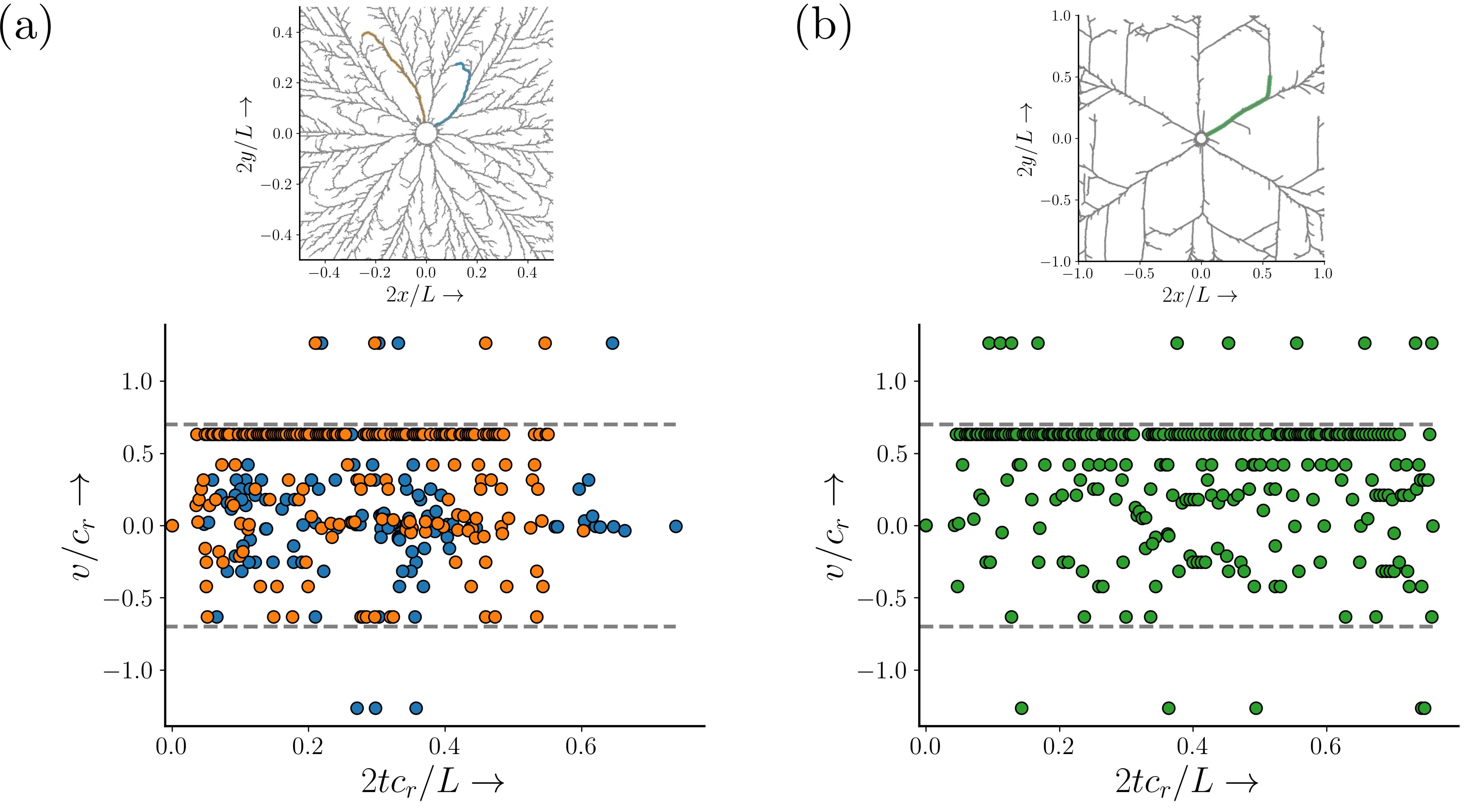}
	\caption{Microscopic crack tip speed $v$ normalized by the Rayleigh wave speed $c_r$. (a) Two distinct chains from Case 2 (blue corresponds to the chain in Fig.~\ref{fig:micro_bond_break}). (b) A selected chain from the low residual stress Case 4 (green).}
	\label{fig:crack_speed}
\end{figure}

A direct consequence of this non-sequential failure is that the apparent microscopic crack speed, $v$, can transiently exceed the Rayleigh wave speed $c_r$, as plotted in Fig.~\ref{fig:crack_speed}. Here, $v$ is calculated as the ratio of the incremental chain length to the time interval between adjacent bond failures. Figure \ref{fig:crack_speed}(a) tracks $v$ for two chains in Case 2 (blue and orange), while Fig. \ref{fig:crack_speed}(b) tracks a chain from the low-stress Case 4 (green). The spikes where $v > c_r$ (and the negative values, which represent a downstream bond breaking before an upstream one) are signatures of the micro-branching events described above. Despite these fluctuations, it is significant that for the majority of the propagation history, the crack speed remains below $0.7c_r$ (dashed line).

These results support a dynamic hypothesis for branching in residually stressed solids: as the crack accelerates toward a critical velocity (typically $\sim 0.7c_r$ \cite{yoffe1951lxxv, freund1998dynamic}), single-tip propagation becomes unstable. This instability manifests as the nucleation of micro-cracks ahead of the tip, altering the local stress field and, under sufficient residual stress, triggering macroscopic bifurcation.

\subsection{Universal scaling of fragmentation in toughened glass}\label{sec:glass_result}
\begin{figure}[h!]
	\centering
	\includegraphics[width = \columnwidth]{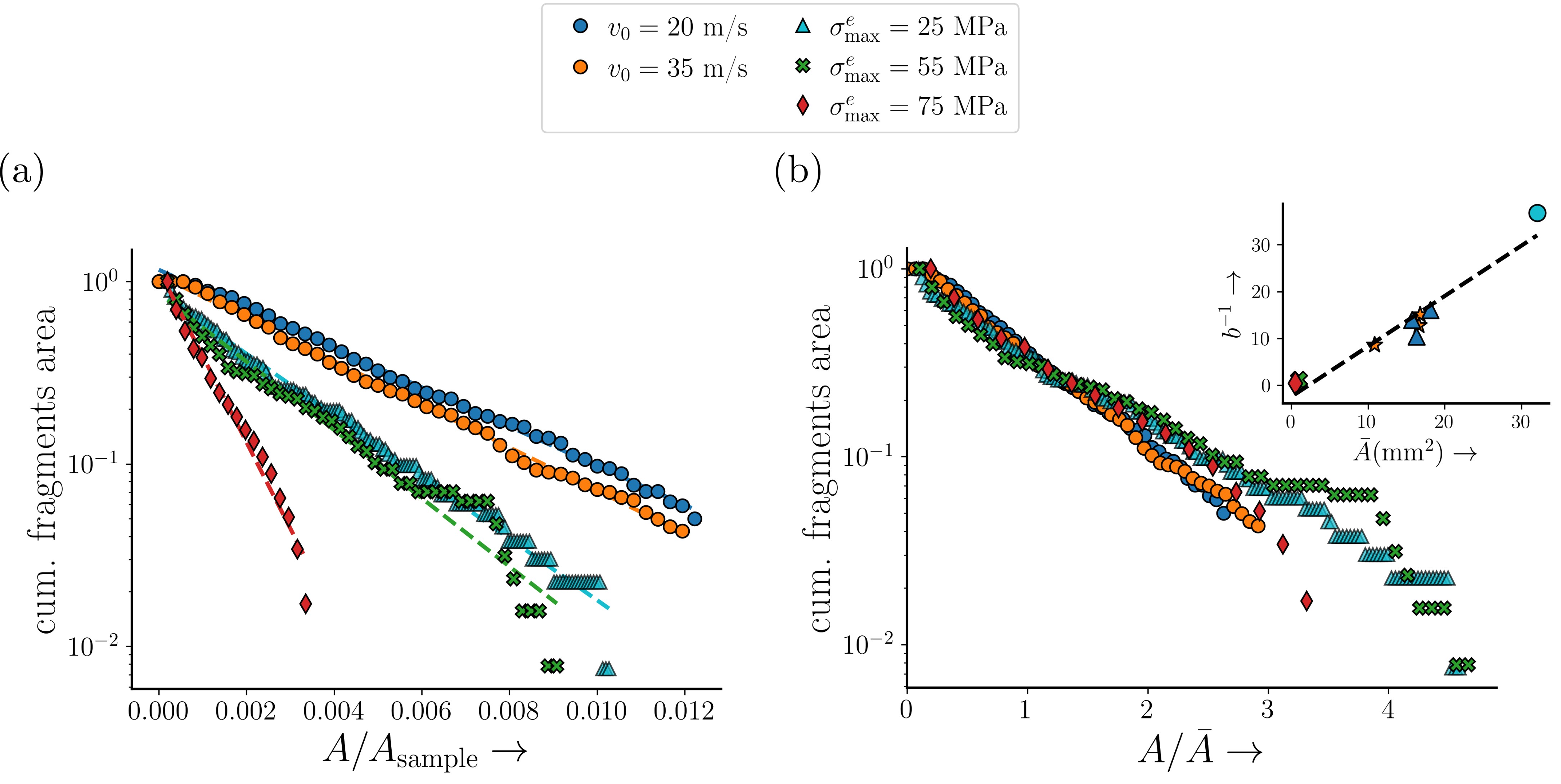}
	\caption{(a) Cumulative fragment area distributions from current and historical experiments. (b) Data collapse obtained by normalizing fragment area $A$ by the mean area $\bar{A}$. Inset: Linear relationship between the decay constant $b^{-1}$ and $\bar{A}$.}
	\label{fig:scaled_cum_exp}
\end{figure}

We finally address the characteristic length scale of fragmentation that emerges in both the experiments and simulations---this is, in fact, a defining feature that distinguishes toughened glass from ordinary brittle solids. We evaluate the statistical nature of the fragmentation process in Figure \ref{fig:scaled_cum_exp}(a), which compares the cumulative fragment area distributions obtained from our experiments against historical benchmarks. Our data, recorded at impact velocities of $v_0 = 20$ m/s (blue \ding{108}) and $v_0 = 35$ m/s (orange \ding{108}), are plotted alongside results from Zijlstra \emph{et al.}\cite{zijlstra1969fracture} (cyan \ding{115}) and Tandon \& Glass \cite{tandon2015fracture} (green \ding{54}, red \ding{169}). These historical datasets provide a robust baseline, encompassing different sample geometries (circular plates) and a wide range of residual stress levels ($25$ to $75$ MPa; see \suppMat\suppref{S3} for details). Despite the diversity in experimental conditions and specimen configurations, all datasets exhibit a remarkably consistent trend: the distributions follow a clear exponential decay, that is very well described by a relationship of the form $y = b \log A + \log a$.

Crucially, normalizing the fragment area $A$ by the mean area $\bar{A}$ reveals a striking universality: all datasets collapse onto a single master curve (Fig. \ref{fig:scaled_cum_exp}(b)). This collapse is not merely statistical; it implies physical self-similarity, demonstrating that the complex fragmentation process is governed by a single characteristic length scale proportional to $\sqrt{\bar{A}}$. The inset in Fig. \ref{fig:scaled_cum_exp}(b) provides quantitative confirmation, establishing a linear relationship between the exponential decay coefficient $b^{-1}$ and the mean area $\bar{A}$ (slope $\approx 1.07$).

\begin{figure}[h!]
	\centering
	\includegraphics[width = 0.6\columnwidth]{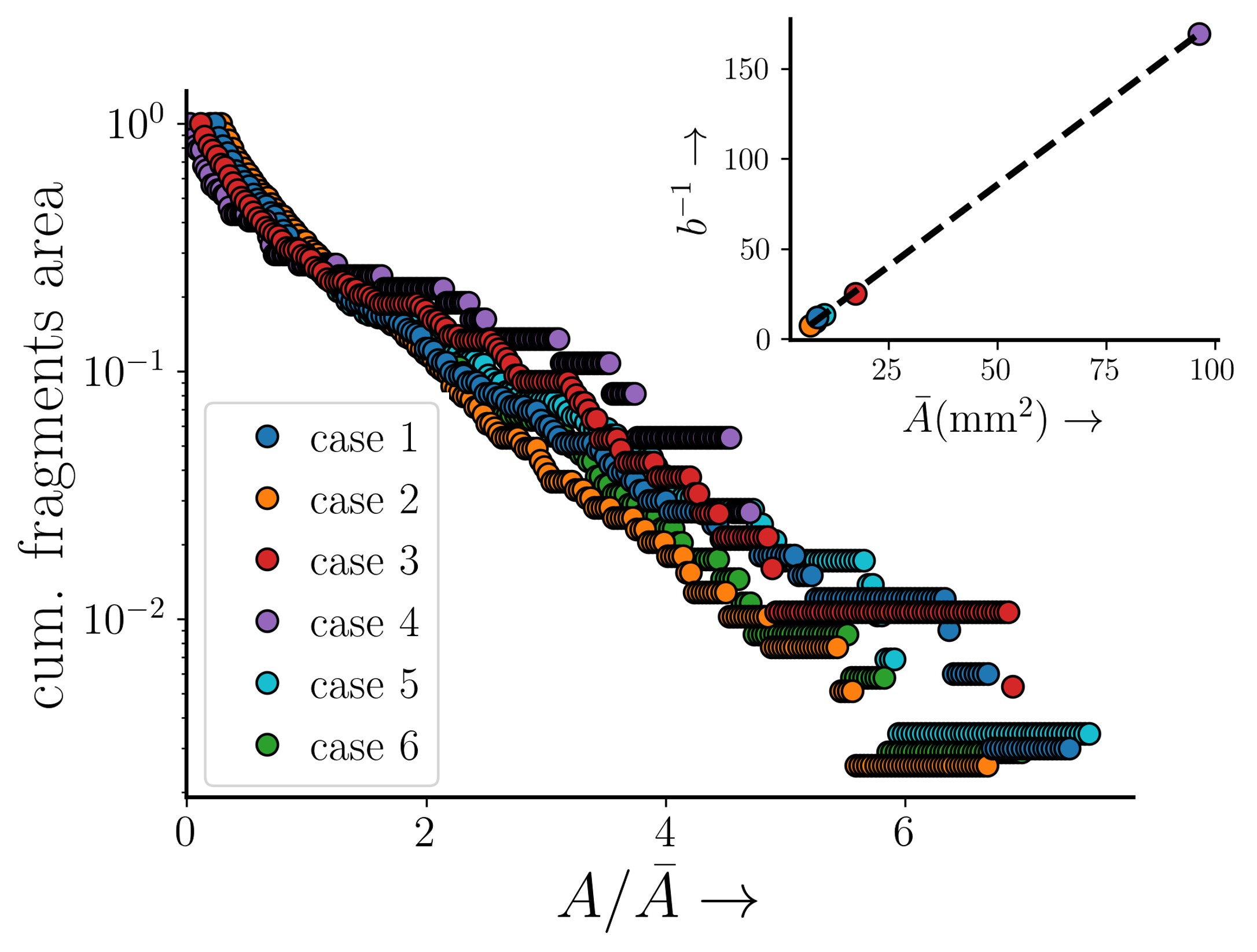}
	\caption{Universal collapse of simulated cumulative fragment area distributions when normalized by $\bar{A}$, across all numerical cases. Inset: Linear scaling between $b^{-1}$ and $\bar{A}$.}
	\label{fig:scaled_cum_num}
\end{figure}

This universality is robustly captured by our numerical framework. As shown in Figure \ref{fig:scaled_cum_num}, the simulated distributions for all cases (Cases 1–6) exhibit an identical data collapse when normalized by $\bar{A}$ (see also \suppMat\suppref{S5}). Although the specific proportionality constant differs slightly between simulation and experiment (slope $\approx 1.8$ vs. $1.07$), the fundamental scaling law remains invariant. This confirms that the model successfully reproduces the essential physics governing the fragmentation hierarchy.

While the existence of a characteristic fragment size in toughened glass is not new \cite{barsom1968fracture, grady2008fragment, vocialta2018numerical}, previous models often relied on idealized single-fragment energy balances to explain their occurrence. Our results explicitly demonstrate that even for complex, chaotic-looking fracture networks generated via dynamic impact, the entire statistical distribution is governed by the average fragment size. Consequently, understanding the complex fragmentation of toughened glass reduces to determining how the scalar quantity $\bar{A}$ scales with impact velocity, residual stress magnitude, and stress profile.

\section{Discussion}
\label{sec:discussion}
We now contextualize our model's assumptions and findings within the broader landscape of experimental and theoretical fracture mechanics, as pertaining to failure of residually stressed brittle solids.

\subsection{Validity of the 2D in-plane approximation}\label{sec:2d_prob}
We have simplified the complex 3D fragmentation of toughened glass into a 2D problem focused on the tensile mid-plane, \emph{cf.} Fig~\ref{fig:dynamic_steps}(a). This reduction is motivated by the observation that fragmentation is initiated within this tensile core and, for thin plates, the final fragment pattern is predominantly two-dimensional. Both our experimental results and recent results in the literature \cite{nielsen2017deformations} confirm that fragments maintain a thickness consistent with the full glass depth (up to 5 mm). This suggests that the in-plane tensile stresses at the mid-plane are the primary drivers of the fracture network, making the 2D approximation reasonable for thin plates ($< 5$ mm).

However, real cracks are inherently 3D objects with finite fronts rather than point-like tips. As a crack propagates, its front advances simultaneously through both the compressive surface layers and the tensile core. High-speed imaging \cite{nielsen2009fracture} has demonstrated that while cracks may propagate at different rates in these zones, the fronts remain coupled. The simplified 2D model cannot capture out-of-plane dynamics, such as the arrest of cracks as they enter the surface compression zone—a mechanism that keeps fragments interlocked. While a full 3D model would be required to capture these effects, the 2D plane-stress approach successfully captures the dominant topological features of the fragmentation network.

\subsection{Justification for the inelastic strain profile}\label{sec:choice_inelastic}
In chemically toughened glass, the compressive surface zone is typically thin ($\sim 400$ $\mu$m), characterized by a stress profile that decays sharply (often following a complementary error function \cite{arakawa2016ultrasonic}) before transitioning into a thicker central tensile region. While the tensile stress near the mid-plane is often assumed to be uniform \cite{tandon2015fracture}, it must vanish at the free edges to satisfy boundary conditions. Our assumed profile captures this essential decay. Although the actual stress variation may arguably be more complex—potentially requiring a superposition of Chebyshev polynomials \cite{balan2015assessment}—the specific polynomial form chosen here is a first-order approximation intended to capture the essential physics of the gradient.

Despite this simplification, the model's predictions are qualitatively consistent with experimental reality. Both simulations and projectile-impact experiments exhibit the same transition from coarse to fine fracture patterns with increasing energy, and both adhere to an exponential fragment area distribution.
Two significant challenges remain for a more precise quantitative reconstruction. First, experimentally measuring the residual stress profile deep within the mid-plane is non-trivial \cite{arakawa2016ultrasonic, efferz2024photoelastic, tandon1990residual}, often requiring reliance on 3D finite-element estimations \cite{soules1987finite, nielsen2010finite, pourmoghaddam2016numerical, pourmoghaddam2018finite}. Second, our model inputs inelastic strain (eigenstrain) rather than stress directly. The inverse problem of converting a measured residual stress field into an eigenstrain field is mathematically complex \cite{garfinkel1970ion, balan2015assessment, korsunsky2009eigenstrain, achintha2011eigenstrain}. Addressing these challenges in future work would allow for a direct validation of the assumed strain profile.

\subsection{Microscopic crack dynamics and branching mechanisms}
Our observation of microscopic crack speeds transiently exceeding the Rayleigh wave speed $c_r$ mirrors the well-known Burridge-Andrews mechanism \cite{xia2004laboratory, andrews1976rupture}. Originally observed in shear cracks along bimaterial interfaces, this mechanism allows apparent "supershear" speeds when daughter cracks nucleate ahead of the main tip and coalesce retroactively. It is plausible that a similar \lq process zone\rq\ mechanism operates at the microscopic scale in toughened glass, where the high residual tension facilitates damage nucleation ahead of the crack tip.

Physically, as crack speeds approach $c_r$, the fracture surface transitions from a smooth mirror/mist region to a rough hackle region \cite{johnson1966shape, kirchner1987criteria, lubomirsky2024quenched}. This roughness is a signature of instability and incipient branching. In our simulations, we observe numerous micro-branches that initiate but fail to develop into macroscopic cracks. This behavior is consistent with 3D peridynamic simulations \cite{stewart2023modeling} and experimental fractography of ion-exchanged glass, which reveals "tongue-like" features \cite{glass2007characterization, quinn2009history} as precursors to branching. These fracture tongues represent local bifurcations that may not reach the criticality required for full macroscopic branching. We propose that the "arrested micro-branches" observed in our network model are the numerical analogues of these physical fracture tongues.

\section{Summary and Conclusions}\label{sec:conclusions}
This study investigated the fundamental mechanisms of dynamic fragmentation in residually stressed glass, focusing on the determinants of fragment size, fracture patterns, and the underlying crack dynamics—aspects that have remained incompletely understood.

We first reported on high-velocity impact tests on commercial ion-exchanged glass using a sharp steel dart. The experiments revealed that, while the fracture pattern shifts from coarse to fine with increasing impact velocity, the cumulative fragment area distribution consistently follows an exponential decay. In order to better capture the underlying physics numerically, we introduced a bond-network framework that explicitly incorporates residual stress via an inelastic/eigenstrain strain distribution, $\boldsymbol{\epsilon}^*(x,y)$. We demonstrated that the mechanical effect of residual stress can be rigorously modeled by applying an equivalent body force $\mathbf{f}_i^\text{inelastic}$ (Eqs. \ref{eq:inelastic_bond_force} and \ref{eq:inelastic_force}) and a corresponding surface traction $\mathbf{T}^*\mathbf{\hat{n}}$. The model requires two primary inputs: the specific inelastic strain profile—assumed here to be axisymmetric (Eq. \ref{eq:inelastic_strain})—and a breaking criterion ($\epsilon_{ij}^\text{max} \geq \epsilon_b$) formulated to ensure mesh-independent crack paths.

Qualitative validation was achieved by comparing simulated fracture patterns under varying radial loading rates $\dot{u}_r$ against experimental observations. The model successfully reproduces key experimental features:
\begin{itemize}
    \item The transition from coarse to fine fragmentation with increasing loading rate.
    \item The persistence of the exponential size distribution across different rates.
    \item The formation of "flower-like" fragments near the impact center—a region often obscured in experiments but clearly resolved in our simulations, consistent with literature reports \cite{tandon2015fracture}.
\end{itemize}
Furthermore, the model elucidates the sequential dynamics of fragmentation: primary cracks nucleate and propagate radially before bifurcating into secondary branches, which subsequently merge to isolate fragments.

Exploiting the model's predictive capability, we analyzed the distinct roles of the residual stress magnitude (degree of tempering) and the stress gradient profile. Our results show that finer fragmentation can be achieved via two distinct pathways: increasing the peak residual stress or steepening the stress gradient. This implies that a glass with lower overall stored energy but a sharper stress profile can yield fragmentation performance comparable to highly tempered glass.

Finally, at the microscopic scale, our analysis of bond-breaking dynamics reveals that:
\begin{enumerate}
    \item Local crack tip speeds generally remain below $0.7c_r$, but can transiently exceed the Rayleigh wave speed $c_r$. This apparent supershear behavior arises from a mechanism analogous to the Burridge-Andrews effect \cite{xia2004laboratory, andrews1976rupture}, where damage nucleates ahead of the main tip and coalesces retroactively.
    \item Macroscopic branching is not an instantaneous event but the result of multiple "attempted" micro-branches. The survival of these micro-branches depends on local dynamic stress interactions. We propose that these arrested micro-branches correspond to the "tongue-like" features observed on fracture surfaces in post-mortem fractography.
\end{enumerate}

Collectively, these findings provide a robust framework for predicting fragmentation in toughened glass, linking the macroscopic statistical laws—specifically the universal scaling of average fragment size—to the microscopic physics of dynamic crack branching.

\bibliography{ref_bib}
\bibliographystyle{vancouver}

\end{document}